 \documentclass{llncs} 
\newtheorem{defn}{Definition} 

\newtheorem{thm}{Theorem}
\newtheorem{lem}[defn]{Lemma}

\newtheorem{rem}[defn]{Remark}
\begin{document}

\title{An Algorithm for Road Coloring}
\author{A.N. Trahtman\thanks{Email: trakht@macs.biu.ac.il}}

\institute{Bar-Ilan University, Dep. of Math., 52900, Ramat Gan, Israel}

\maketitle

\begin{abstract}
A coloring of edges of a finite directed graph turns the graph into
a finite-state automaton.
The synchronizing word of a deterministic automaton is a word in
the alphabet of colors (considered as letters) of its edges that
maps the automaton to a single state.  A coloring of edges of a
directed graph of uniform outdegree (constant outdegree of any vertex)
 is synchronizing if the coloring turns the graph
into a deterministic finite automaton possessing a synchronizing word.

The road coloring problem is the problem of synchronizing coloring
of a directed finite strongly connected graph of uniform outdegree
if the greatest common divisor of the lengths of all its cycles is
one. The problem posed in 1970 has evoked noticeable interest
among the specialists in the theory of graphs, automata, codes,
symbolic dynamics as well as among the wide mathematical community.

A polynomial time algorithm of $O(n^3)$ complexity in the
worst case and quadratic in the majority of studied cases for the road
coloring of the considered graph is presented below. The work is
based on the recent positive solution of the road coloring problem.
The algorithm was implemented in the freeware package TESTAS.
\end{abstract}

 \section*{Introduction}
The road coloring problem was stated almost 40 years ago
\cite{AW}, \cite {AGW}  for a strongly connected directed  finite
deterministic graph of uniform outdegree where the greatest common
divisor (gcd) of the lengths of all its cycles is one.
The edges of the graph being unlabelled, the task is to find a
labelling of the edges that turns the graph into a deterministic
finite automaton possessing a synchronizing word.
The outdegree of the vertex can be considered also as the size of an
alphabet where the letters denote colors.

The condition on gcd is necessary \cite{AGW}, \cite{CKK}.
 It can be replaced by the equivalent property that there does not exist
 a partition of the set of vertices on subsets $V_1$, $V_2$, ...,
 $V_k = V_1$ ($k>2$) such that every edge which begins in $V_i$ has its end
in $V_{i+1}$ \cite {CKK}, \cite {OB}.

Together with the \v{C}erny conjecture \cite {Ce}, \cite {CPR}, \cite {IS},
 \cite {ID}, \cite{MS}, \cite{Ry} the road coloring problem used to belong
to the most fascinating problems in the theory of finite automata.
The popular Internet Encyclopedia "Wikipedia" mentions it on the list of the most
interesting unsolved problems in mathematics.

For some results in this area, see \cite{BM},
\cite {Ca}, \cite {Fi}, \cite {GKS}, \cite{HJ}, \cite {JS}, \cite {Ka},
\cite {OB}, \cite {PS}.
 A detailed history of investigations can be found in \cite {Ca}.
The final positive solution of the problem is stated in \cite{Ti}.

An algorithm for road coloring oriented on DNA computing \cite{JK}
is based on the massive parallel computing of sequences of length
$O(n^3)$. The implementation of the algorithm as well as the
implementation of effective DNA computing is still an open problem.

Another new algorithm for road coloring (ArXiv \cite{BP}) as well as our algorithm
 below is based on the proof of \cite {Ti}. This proof is constructive and leads
to an algorithm that finds a synchronized labelling with cubic
worst-case time complexity. Both of the above mentioned algorithms use
concepts and ideas of the considered proof together with the
concepts from \cite{CKK}, \cite{Ka}, but use different methods
to reduce the time complexity. A skillful study of the graph was added
in \cite{BP}. Their algorithm was declared as quadratic, but some
uncertainties in the proofs must be removed before publication and
implementation.

Our algorithm for the road coloring (ArXiv \cite {Tv}) reduces the time complexity
with the help of the study of two cycles with common vertex (Lemma \ref{al}). It
gives us the possibility to reduce quite often the time complexity.

The theorems and lemmas from \cite {Ti} and \cite {Tr} are presented below
without proof. The proofs are given only for new (or modified) results.
The time complexity of the algorithm for a graph with $n$
vertices and $d$ outgoing edges of any vertex is $O(n^3d)$ in the
worst case and quadratic in the majority of the studied cases.
The space complexity is quadratic. At present, only this algorithm
for road coloring is implemented.

 The description of the algorithm is presented below together with some pseudo codes of the
implemented subroutines. The algorithm is implemented in the
freeware package TESTAS (http://www.cs.biu.ac.il/$\sim$trakht/syn.html) \cite{TBC}.
The easy access to the package ensures the possibility to everybody to verify
the considered algorithm.

The role of the road coloring and the algorithm is substantial
also in education. "The Road Coloring Conjecture makes a nice supplement
to any discrete mathematics course" \cite{Ra}. The realization of the algorithm
is demonstrated on the basis of a linear visualization program \cite{BCT}
and can analyze any kind of input graph.

 \section*{Preliminaries}
As usual, we regard a directed graph with letters assigned to its edges as a finite
automaton, whose input alphabet $\Sigma$ consists of these letters. The graph is called
{\it a transition graph} of the automaton. The letters from $\Sigma$ can be considered as
colors and the assigning of colors to edges will be called {\it coloring}.

A finite directed strongly connected graph with constant
outdegree of all its vertices where the gcd of lengths of all
 its cycles is one will be called an {\it $AGW$ graph} (as introduced by Adler,
Goodwyn and Weiss).

 We denote by $|P|$ the size of the subset $P$ of states of an automaton
  (of vertices of a graph).

If there exists a path in an automaton from the state $\bf p$ to
the state $\bf q$ and the edges of the path are consecutively
labelled by $\sigma_1, ..., \sigma_k$, then for
$s=\sigma_1...\sigma_k \in \Sigma^+$ we shall write  ${\bf q}={\bf p}s$.

Let $Ps$ be the set of states ${\bf p}s$ for ${\bf p} \in P$,
 $s \in \Sigma^+$. For the transition graph
$\Gamma$ of an automaton, let $\Gamma s$
denote the map of the set of states of the automaton.

 A word $s \in \Sigma^+ $ is called a {\it synchronizing}
word of the automaton with transition graph $\Gamma$
if $|\Gamma s|=1$.

 A coloring of a directed finite graph is {\it synchronizing} if the
coloring turns the graph into a deterministic finite automaton
possessing a synchronizing word.

Bold letters will denote the vertices of a graph and the states
of an automaton.

  A pair of distinct states $\bf p, q$ of an automaton (of vertices of
  the transition graph)  will be called  {\it synchronizing} if
  ${\bf p}s = {\bf q}s$ for some $s \in \Sigma^+$.
In the opposite case, if ${\bf p}s \neq {\bf q}s$ for any $s$, we
call the pair a {\it deadlock}.

A synchronizing pair of states $\bf p$, $\bf q$ of an automaton is
called {\it stable} if for any word $u$ the pair ${\bf p}u, {\bf
q}u$ is also synchronizing \cite{CKK}, \cite{Ka}.

We call the set of all outgoing edges of a vertex a {\it bunch} if
 all these edges are incoming edges of only one vertex.

The subset of states (of vertices of the transition
graph $\Gamma$) of maximal size such that every pair of states
from the set is a deadlock will be called an {\it $F$-clique}.

\section{Some properties of $F$-cliques and stable pairs}
The road coloring problem was formulated for $AGW$ graphs
\cite{AGW} and only such graphs are considered in Sections $1$ and $2$.

Let us recall that a binary relation $\rho$ on the set of the
states of an automaton is called {\it congruence} if $\rho$ is
equivalence and for any word $u$ from ${\bf p}$ $\rho$ ${\bf q}$
follows ${\bf p}u$ $\rho$ ${\bf q}u$.
Let us formulate an important result from \cite{CKK}, \cite{Ka}
in the following form:

 \begin{thm}  $\label {ck}$ \cite{Ka}
Let us consider a coloring of an $AGW$ graph $\Gamma$.
Let $\rho$ be the transitive and reflexive closure of the stability relation
on the obtained automaton.
Then $\rho$ is a congruence relation, $\Gamma/\rho$ is also an
$AGW$ graph and a synchronizing coloring of $\Gamma/\rho$ implies
a synchronizing recoloring of $\Gamma$.
 \end{thm}
 \begin{lem}  $\label {f3}$ \emph{\cite{Ti}, \cite{CKK}}
Let $F$ be an $F$-clique of some coloring of an $AGW$ graph $\Gamma$.
For any word $s$ the set $Fs$ is also an $F$-clique and any state
$\bf p$ belongs to some $F$-clique.
 \end{lem}
  \begin{lem}  $\label {f6}$
Let $A$ and $B$ (with $|A|>1$) be distinct $F$-cliques of
some coloring  of an $AGW$ graph $\Gamma$ such that $|A| -|A \cap B| =1$.
 Then for all ${\bf p} \in A \setminus A \cap B$ and
${\bf q} \in B \setminus A \cap B$, the pair $({\bf p}, {\bf q})$ is stable.
 \end{lem}
Proof.
By the definition of an $F$-clique, $|A|=|B|$ and $|B| -|A \cap B|=1$, too.
 If the pair of states ${\bf p} \in A \setminus B$ and
${\bf q} \in B \setminus A$ is not stable, then for some word
$s$ the pair $({\bf p}s, {\bf q}s)$ is a deadlock.
Any pair of states from the $F$-clique $A$ and from the $F$-clique $B$,
as well as from the $F$-cliques $As$ and $Bs$, is a deadlock. So any pair
of states from the set $(A \cup B)s$ is a deadlock. One has
$|(A \cup B)s| = |As| + 1 = |A| + 1 > |A|$. So the size of the set
$(A \cup B)s$ of deadlocks is greater than the maximal size of
$F$-clique. Contradiction.
\begin{lem}  $\label {f7}$
 If some vertex of an $AGW$ graph $\Gamma$ has two incoming bunches, then
the origins of the bunches form a stable pair by any coloring.
 \end{lem}
Proof. If a vertex ${\bf p}$ has two incoming bunches from ${\bf
q}$ and ${\bf r}$, then the couple ${\bf q}$, ${\bf r}$ is stable
for any coloring because ${\bf q}\sigma ={\bf r}\sigma =\bf p$ for
any $\sigma \in \Sigma$.
\section{The spanning subgraph of an $AGW$ graph}
\begin{defn}  $\label {d1}$
Let us call a subgraph $S$ of an $AGW$ graph $\Gamma$, a {\emph
spanning subgraph} of $\Gamma$, if $S$ contains all vertices of
$\Gamma$ and if each vertex has exactly one outgoing edge. (In
usual graph-theoretic terms it is a 1-outregular spanning subgraph).

 A maximal subtree of a spanning subgraph $S$ with its root on a cycle from $S$ and
having no common edges with the cycles of $S$ is called a {\emph tree} of $S$.

 The length of a path from a vertex ${\bf p}$ through the edges of the tree of the spanning
set $S$ to the root of the tree is called a {\emph  level} of ${\bf p}$ in $S$.

A tree with a vertex of maximal level is called a {\emph maximal tree}.
 \end{defn}
\begin{rem}
Any spanning subgraph $S$ consists of disjoint cycles
and trees with roots on the cycles. Any tree and cycle of $S$ is defined
identically. The level of the vertices belonging to some cycle is zero.
The vertices of the trees except the roots have positive level. The
vertices of maximal positive level have no incoming edge in $S$.
The edges labelled by a given color defined by any coloring form
a spanning subgraph. Conversely, for each spanning subgraph, there exists
a coloring and a color such that the set of edges labelled with this
color corresponds to this spanning subgraph.
\end{rem}
\begin{picture}(50,80)
\end{picture}
\begin{picture}(150,80)
\put(87,54){\vector(0,-1){17}}
\put(87,54){\circle{4}}
\put(105,72){\vector(-1,-1){18}}
\put(105,72){\circle{4}}
\put(69,54){\vector(1,0){18}}
\put(69,54){\circle{4}}
\put(51,72){\vector(1,-1){18}}
\put(51,72){\circle{4}}
\put(87,36){\vector(1,-1){13}}
\put(87,36){\circle{4}}
\put(74,23){\vector(1,1){13}}
\put(74,23){\circle{4}}
\put(87,10){\vector(-1,1){13}}
\put(87,10){\circle{4}}
\put(100,23){\vector(-1,-1){13}}
\put(100,23){\circle{4}}
\put(27,76){max level 3}
\put(40,18){level 0}
\put(94,9){Cycle}
\put(96,76){level 2}
\put(96,51){level 1}
\put(94,32){level 0}
\put(60,44){Tree}
\put(69,41){\vector(3,-1){17}}
\put(69,41){\circle{4}}
 \end{picture}
  \begin{lem}  $\label {p1}$ \emph{\cite{Ti}} \emph{\cite{Tr}}
Let $N$ be a set of vertices of maximal level in some tree of the
spanning subgraph $S$ of an $AGW$ graph $\Gamma$.
Then, via a coloring of $\Gamma$ such that all edges of $S$ have
the same color $\alpha$, for any $F$-clique $F$ holds $|F \cap N|
\leq 1$.
 \end{lem}
\begin{lem}  $\label {f8}$ \emph{\cite{Ti}}
Let $\Gamma$ be an $AGW$ graph with a spanning subgraph
$R$ which is a union of cycles (without trees).
 Then the non-trivial graph $\Gamma$ has another spanning subgraph with exactly
one maximal tree.
\end{lem}
 \begin{lem}  $\label {f9}$
Let $R$ be a spanning subgraph of an $AGW$ graph $\Gamma$. Let $T$ be a maximal tree of $R$ with
 a vertex $\bf p$ of maximal positive level $L$ and with a root $\bf r$ on
a cycle $H$ of $R$.
Let us change the spanning subgraph by means of the following flips:

1)an edge $\bar{a}={\bf a} \to {\bf p}$ replaces the edge $\bar{d}={\bf a} \to {\bf d}$
of $R$ for appropriate vertices ${\bf a}$ and ${\bf d}\neq {\bf p}$,

2) replacing edge $\bar{b}={\bf b} \to {\bf r}$ of $T$ by an edge ${\bf b} \to {\bf x}$
 for appropriate vertices  ${\bf b}$ and ${\bf x} \neq {\bf r}$,

3) replacing edge $\bar{c}={\bf c} \to {\bf r}$ of $H$ by an edge ${\bf c} \to {\bf x}$
for appropriate vertices  ${\bf c}$ and ${\bf x} \neq {\bf r}$.

Suppose that one or two consecutive flips do not increase the number of edges in cycles
(Condition$^*$) and no vertex of $\Gamma$ has two incoming bunches (Condition$^{**}$).
 Then there exists a spanning subgraph with a single maximal non-trivial tree.
 \end{lem}
Proof. In view of Lemma \ref{f8}, suppose that $R$ has non-trivial trees.
Further consideration is necessary only if the maximal tree $T$ is not single.

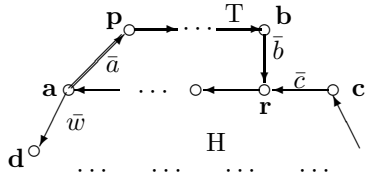
\begin{picture}(100,80)
\end{picture}
\begin{picture}(150,80)
\multiput(61,33)(26,0){3}{\circle{4}}
\put(87,54){\vector(0,-1){18}}
\put(87,56){\circle{4}}
\put(36,56){\circle{4}}
\put(0,10){\circle{4}}

\multiput(85,33)(26,0){2}{\vector(-1,0){21}}
\put(123,11){\vector(-1,2){10}}
\put(12,32){\vector(-1,-2){10}}

\put(13,33){\circle{4}}
\put(27,58){$\bf p$}
\put(85,24){$\bf r$}
\put(3,31){$\bf a$}
\put(-10,4){${\bf d}$}

\put(120,31){$\bf c$}
\put(91,58){$\bf b$}

\put(38,56){\vector(1,0){16}}
\put(69,56){\vector(1,0){17}}

\put(56,54){$\cdots$}
\put(39,30){$\cdots$}
\multiput(16,0)(28,0){4}{$\cdots$}
\put(32,33){\vector(-1,0){17}}
\put(27,40){$\bar{a}$}
\put(12,18){$\bar{w}$}
\put(98,34){$\bar{c}$}
\put(90,45){$\bar{b}$}
\put(65,10){H}
\put(72,59){T}
\put(14,34){\vector(1,1){20}}
\put(13,34){\line(1,1){20}}
 \end{picture}

Our aim is to increase the maximal level $L$ using the three
aforesaid flips. If one of the flips does not succeed, let us go
to the next, assuming the situation in which the previous fails,
and excluding the successfully studied cases. We check at most two
flips together.
 Let us begin from

the edge $\bar {a}$)  Suppose first ${\bf a} \not\in H$. If ${\bf a}$ belongs to the path in $T$
 from ${\bf p}$ to ${\bf r}$ then a new cycle with part of the path and the edge
${\bf a} \to {\bf p}$ is added to $R$ extending the number of vertices in its
cycles in spite of Condition$^*$ of lemma. In the opposite case the level of ${\bf a}$ is $L+1$
in a single maximal tree.

So let us assume ${\bf a} \in H$.
In this case the vertices ${\bf p}$, ${\bf r}$ and ${\bf a}$ belong to
a cycle $H_1$ of a new spanning subgraph $R_1$ obtained by removing $\bar{d}$
and adding $\bar{a}$. So we have the cycle $H_1 \in R_1$ instead of $H \in R$.
If the length of the path from ${\bf r}$ to ${\bf a}$
in $H$ is $r_1$, then $H_1$ has length $L+r_1+1$. A path from ${\bf r}$
to ${\bf d}$ of the cycle $H$ remains in $R_1$.
Suppose that its length is $r_2$. So the length of the cycle $H$
is $r_1+r_2+1$. The length of the cycle $H_1$ is not greater than
the length of $H$ in view of Condition$^*$.
 So $r_1+r_2+1 \geq L+r_1+1$, whence $r_2
\geq L$. If $r_2 > L$, then the length $r_2$ of the path from
${\bf d}$ to $\bf r$ in a tree of $R_1$ (as well as the level of ${\bf d}$)
 is greater than $L$. The tree containing ${\bf d}$ is the desired single maximal tree.

 So we can assume for further consideration that $L=r_2$ and
 ${\bf a} \in H$. An analogous statement can be stated for any maximal tree.

the edge $\bar {b}$) Suppose that the set of outgoing edges of the vertex ${\bf b}$ is
not a bunch. So one can replace in $R$ the edge $\bar{b}$
by an edge $\bar{v}={\bf b} \to {\bf v}$
(${\bf v}\neq {\bf r}$).

The vertex ${\bf v}$ could not belong to $T$ because in this case a new cycle
is added to $R$ in spite of Condition$^*$.

If the vertex ${\bf v}$ belongs to another tree of $R$ but not to the cycle $H$,
then $T$ is a part of a new tree $T_1$ with a new root of a new spanning
subgraph $R_1$ and the path from ${\bf p}$ to the new root has a length
greater than $L$. Therefore the tree $T_1$ is the unique maximal tree in $R_1$.

If ${\bf v}$ belongs to some cycle $H_2 \neq H$ in $R$, then
together with replacing $\bar{b}$ by $\bar{v}$, we also replace the
 edge $\bar{d}$ by $\bar{a}$. So we extend the path from ${\bf p}$
to the new root ${\bf v}$ of $H_2$ at least by the edge
$\bar{a}={\bf a} \to {\bf p}$ and there is a unique maximal tree of level $L_1>L$
which contains the vertex $\bf d$.

Now it remains only the case when ${\bf v}$ belongs to the cycle
$H$. The vertex ${\bf p}$ also has level $L$ in a new tree $T_1$
with root ${\bf v}$. The only difference between $T$ and $T_1$
(just as between $R$ and $R_1$) is the root and the incoming edge
of this root. The new spanning subgraph $R_1$ has the same number of
vertices in their cycles just as does $R$. Let $r_2^{'}$ be the length
of the path from ${\bf d}$ to ${\bf v} \in H$.

For the spanning subgraph $R_1$, one can obtain $L=r_2^{'}$ just as it
was done earlier in the case of the edge $\bar {a}$) for $R$.
From  ${\bf v} \neq {\bf r}$ follows $r_2^{'}  \neq r_2$,
though $L=r_2^{'}$ and $L=r_2$.

So for further consideration suppose that the set of outgoing
edges of the vertex ${\bf b}$ is a bunch to ${\bf r}$.

The edge $\bar {c}$) The set of outgoing edges of the vertex ${\bf c}$
 is not a bunch in virtue of Condition$^{**}$ (${\bf r}$ has another
 bunch from ${\bf b}$.)

Let us replace in $R$ the edge $\bar{c}$ by an edge $\bar{u}={\bf
c} \to {\bf u}$ such that ${\bf u}\neq {\bf r}$. The vertex ${\bf
u}$ could not belong to the tree $T$ because one has in this case
a cycle with all vertices from $H$ and some vertices of $T$ whence
its length is greater than $|H|$ and so the number of vertices in
the cycles of a new spanning subgraph grows in spite of Condition$^*$.

If the vertex ${\bf u}$ does not belong to $T$, then the tree $T$
is a part of a new tree with a new root. The path from ${\bf p}$ to
the new root is extended at least by a part of $H$ starting at the former
root $\bf r$. The new level of ${\bf p}$ therefore is maximal and
greater than the level of any vertex in another tree.

Thus in any case we obtain a spanning subgraph with a single non-trivial
maximal tree.
\begin{lem}  $\label {t1}$
For some coloring of any $AGW$ graph $\Gamma$, there exists a stable pair of states.
  \end{lem}
Proof.
We exclude the case of two incoming bunches of a vertex in
virtue of Lemma \ref{f7}. There exists a coloring such that for some color
$\alpha$, the corresponding spanning subgraph $R$ has maximum edges in cycles.

By Lemma \ref{f9}, we must consider now a spanning subgraph $R$ with a single
 maximal tree $T$. Let the root $\bf r$ of $T$ belong to the cycle $C$.

By Lemma \ref{f3}, in a strongly connected transition graph for
every word $s$ and $F$-clique $F$ of size $|F| > 1$, the set $Fs$
also is an $F$-clique of the same size and for
any state $\bf p$ there exists an $F$-clique $F$ such that ${\bf
p} \in F$.

In particular, some $F$-clique $F$ has a non-empty intersection with the set
$N$ of vertices of maximal level $L$. The set $N$ belongs to one
tree, whence by Lemma \ref{p1} $|N \cap F|=1$. Let ${\bf p} \in N \cap F$.

 The word $\alpha^{L-1}$ maps $F$ on an $F$-clique $F_1$ of size
$|F|$. One has $|F_1 \setminus C|=1$ because any sequence of length
$L-1$ of edges of color $\alpha$ in any tree of $R$ leads to a cycle.
For the set $N$ of vertices of maximal level,
 $N\alpha^{L-1} \not\subseteq C$ holds. So
$|N\alpha^{L-1}  \cap F_1|=|F_1 \setminus C|=1$,
${\bf p}\alpha^{L-1}  \in F_1 \setminus C$
 and $|C \cap F_1|=|F_1|-1$.

Let the integer $m$ be a common multiple of the lengths
of all considered cycles colored by $\alpha$.
 So for any $\bf r$ in $C$ as well as in $F_1 \cap C$
holds ${\bf r}\alpha^m={\bf r}$. Let $F_2$ be $F_1\alpha^m$.
We have $F_2 \subseteq C$ and $C \cap F_1 =F_1 \cap F_2$.

Thus the two $F$-cliques $F_1$ and $F_2$ of size $|F_1|>1$ have $|F_1|-1$ common
vertices. So $|F_1 \setminus (F_1 \cap F_2)|=1$, whence by Lemma \ref{f6},
the pair of states $\bf p\alpha^{L-1}$ from $F_1 \setminus (F_1 \cap F_2)$
and $\bf q$ from $F_2 \setminus (F_1 \cap F_2)$ is stable. It is obvious that
${\bf q}= \bf p\alpha^{L+m-1}$.
\begin{thm}  $\label {t}$ \emph{\cite{Ti}}
Every $AGW$ graph has a synchronizing coloring.
 \end{thm}
\begin{thm}  $\label {t2}$ \emph{\cite{Tc}}
Let every vertex of a strongly connected directed graph $\Gamma$ have the
same number of outgoing edges. Then $\Gamma$ has synchronizing coloring
if and only if the greatest common divisor of lengths of all its cycles
is one.
 \end{thm}
The goal of the following lemma is to reduce the complexity of the algorithm.
\begin{lem}  $\label {al}$
Let $\Gamma$ be an $AGW$ graph having two cycles $C_u$ and $C_v$. Suppose that
 either $C_u \cap C_v =\{ {\bf p}_1\}$ or $C_u \cap C_v =\{{\bf p}_k$,..., ${\bf p}_1 \}$,
 where all incoming edges of ${\bf p}_i$ develop a bunch from ${\bf p}_{i+1}$ ($i<k$).

 Let $u \in C_u$ and $v \in C_v$ be the distinct edges of the cycles
$C_u$ and $C_v$ leaving ${\bf p}_1$.
Let $R_u$ be a spanning subgraph with all edges from $C_u$ and $C_v$ except
$u$. The spanning subgraph $R_v$ is obtained from $R_u$ by removing $v$ and
adding $u$.

Then at least one of two spanning subgraphs $R_u$, $R_v$ has a unique maximal
tree whose root is ${\bf p}_1$.
  \end{lem}
Proof.
 Let us add to $R_u$ the edge $u$ and consider a set of trees with roots on
the cycles $C_u$ and $C_v$. The trees have no common vertices and have no vertices
except a root on the cycles $C_u$ and $C_v$. The same set of trees can be obtained
by adding the edge $v$ to $R_v$.

Let us define the levels of vertices of a tree as in the case of a spanning subgraph
and consider the set of maximal trees (the trees with a maximal vertex level).

If all maximal trees have a common root, then $R_u$ (and also $R_v$)
is a spanning subgraph with a unique maximal tree.

If maximal trees have different roots, then let as take a maximal tree $T$ with
root ${\bf r}$ such that the length of the path $P$  from ${\bf r}$ to ${\bf p}_1$
on the cycle $C_u$ (or $C_v$) is maximal.
If $P$ belongs to $C_u$, then the tree $T$ is extended by the path $P$, whence
$R_u$ has a unique maximal tree. In the opposite case, $R_v$ has a unique maximal tree.

\section{The algorithm for synchronizing coloring}

Let us start with transition graph of an arbitrary deterministic complete finite automaton.
\subsection{Preliminary steps}

The study is based on Theorem \ref{t2}. A synchronizing graph has a sink strongly
connected component ($SCC$).  Our aim is to reduce the study to
sink $SCC$ (if it exists) in order to remove non-synchronizing graphs
without sink $SCC$ and then check the condition on $gcd$.

 The function \large CheckSinkSCC \normalsize verifies the existence of sink $SCC$.
We use the linear algorithm for finding strongly connected
components $SCC$ \cite{AHU}, \cite{Ta}.

Then we remove all $SCC$ as having outgoing edges to other $SCC$.
If only one $SCC$ remains then let us continue. In the opposite
case a synchronizing coloring does not exist.

We study a strongly connected graph (with one $SCC$).
The function \large FindGCDofCycles \normalsize finds the great common divisor
($gcd$) of lengths of cycles of the automaton and verifies the
 necessary conditions of synchronizability ($gcd=1$).

Let $\bf p$ be an arbitrary fixed vertex. Suppose $d({\bf p})=1$.
Then we use a depth-first search from $\bf p$. For an edge ${\bf
r} \to {\bf q}$ where $d({\bf r})$ is already defined and $d({\bf
q})$ is not, suppose $d({\bf q})=d({\bf r})+1$. If $d({\bf q})$ is
defined, let us add the non-zero difference $abs(d({\bf
q})-1-d({\bf r}))$ to the set $D$. The integer from $D$ is a
difference of lengths of two paths from $\bf p$ to $\bf q$. In a
strongly connected graph, the $gcd$ of all elements of $D$ is also
a $gcd$ of lengths of all cycles \cite{AW}, \cite{Tc}.

If $gcd=1$ for all integers from $D$, then the graph has synchronizing coloring.
In opposite case the answer is negative.
So we reduce the investigation to an $AGW$ graph.

Let us proceed with an arbitrary coloring of such a graph $\Gamma$ with
$n$ vertices and constant outdegree $d$.
The considered $d$ colors define $d$ spanning subgraphs of the graph.

We keep the preimages of vertices and colored edges by
any transformation and homomorphism.

If there exists a loop in $\Gamma$ around a state $\bf r$, then let
us color the edges of a tree whose root is $\bf r$ with the same
color as the color of the loop.  The other edges may be colored
arbitrarily. The coloring is synchronizing \cite{AGW}. The
function \large FindLoopColoring \normalsize finds the coloring.

\subsection{Help subroutines}
In the case of two incoming bunches of some vertex, the origins of
these bunches develop a stable pair by any coloring (Lemma
\ref{f7}). We merge both vertices in the homomorphic image of the
graph (Theorem \ref{ck}) and obtain according to the theorem a new
$AGW$ graph of a smaller size. The pseudo code of corresponding
procedure returns two such origins of bunches (a stable pair).

The linear search of two incoming bunches and of the loop can be made at any stage
of the algorithm.

The function \large HomonorphicImage \normalsize of linear
complexity reduces the size of the considered automaton and its
transition graph. The congruence classes of the homomorphism are
defined by a stable pair (Theorem \ref{ck}). A new $AGW$ graph of
a smaller size will be the output.

The main part of the algorithm needs the parameters of the spanning subgraph:
levels of all vertices, the number of vertices (edges) in cycles, trees, next and former vertices.
We keep the tree and the cycle of any vertex, the root of the tree.
 We form the set of vertices of {\it maximal} level and the set of {\it maximal}
trees. The function  \large{FindParameters} \normalsize(spanning subgraph $S$, parameters)
is linear and used by any recoloring step.

The subroutine \large MaximalTreeToStablePair \normalsize of
linear complexity finds a stable pair in a given spanning subgraph
with unique maximal tree. The stable pair consists of two
beginnings of incoming edges of the root of the unique maximal
tree (Lemma \ref {t1}).

\subsection{A possibility to reduce the complexity}

Our algorithm as well as the algorithm of \cite{BP} is based on \cite{Ti}.
Only this section essentially differs in both these papers.

If there are two cycles with one common vertex (path) then we use
Lemma \ref{al} and find a spanning subgraph with single maximal
tree. Then after coloring edges of spanning subgraph by a color
$\alpha$, we find a stable pair (beginnings of two incoming edges
to the root  of the tree).

 The function  \large TwoCyclesWithIntersection  \normalsize
 as a rule returns a pair of cycles with common vertex (path).
The vast majority of digraphs contains such a pair of cycles. The
goal of the subroutine is to omit the cubic complexity of the
algorithm. The search of a stable pair is linear in this case and
thus the whole algorithm is quadratic.

\large{TwoCyclesWithIntersection} \normalsize(graph $G$)

1 levels of all vertices first are negative

2 level(${\bf r})=1$ {\bf and} add ${\bf r}$ to stack

3 {\bf for} every vertex  ${\bf q}$ from stack

4 \quad {\bf do}

5 \qquad {\bf for} every letter $\beta$

6 \qquad \quad  {\bf do}

7 \qquad \qquad add ${\bf q}\beta$ to stack

8 \qquad \qquad {\bf if} level(${\bf q}\beta)\geq 0$

9 \qquad \qquad \quad  level(${\bf q}\beta)=$level(${\bf q})+1$

10 \qquad \qquad \quad  keep the cycle C of vertices ${\bf q}\beta,{\bf q}$ {\bf and break} from both cycles

11 \quad remove ${\bf q}$ from stack

12 {\bf for} every vertex  ${\bf r}$

13 \quad {\bf do}

14 \qquad  {\bf if} ${\bf r} \not\in C$ level((${\bf r})=-1$ (for a search of second cycle)

15 {\bf for} every vertex  ${\bf q}$ from cycle $C$

16 \quad {\bf do}

17 \qquad  ${\bf r}={\bf q}\alpha$

18 \qquad  {\bf for} every letter $\beta$

19 \quad \qquad  {\bf do}

20 \qquad \qquad  {\bf if} ${\bf r}\ne {\bf q}\beta$ {\bf break}

21 \qquad {\bf if} ${\bf r} \ne {\bf q}\beta$ {\bf break}

22 add ${\bf q}$ to stack 1 (possible intersection of two cycles)

23 {\bf for} every vertex  ${\bf r}$ from stack 1

24 \quad {\bf do}

25 \qquad {\bf for} every letter $\beta$

26 \qquad \quad  {\bf do}

27 \qquad \qquad {\bf if} level(${\bf r}\beta)<0$

28 \qquad \qquad \quad  level(${\bf r}\beta)=$level(${\bf r})+1$

29 \qquad \qquad \quad add ${\bf r}\beta$ to stack 1

30 \qquad \qquad  \quad {\bf if} ${\bf r}\beta={\bf q}$ (found second cycle)

31 \qquad \qquad \qquad develop trees with roots on both cycles, find maximal trees

32 \qquad \qquad \qquad color the edge $v$ from $\bf q$ on cycle of maximal tree by color 2

33 \qquad \qquad \qquad  color  the edges of trees and both cycles except $v$ by color 1

 34 \qquad \qquad \qquad \large{FindParameters} \normalsize(spanning subgraph of color 1)

35 \qquad \qquad \qquad \large{MaximalTreeToStablePair} \normalsize (subgraph, $\bf p$, $\bf s$)

36 \qquad \qquad \qquad {\bf return} $\bf p$, $\bf s$ (stable pair)

37 \qquad remove ${\bf r}$ from stack 1

38 {\bf return} False

\subsection{The recoloring of the edges}
A repainting of the edges of the transition graph for to obtain a
spanning subgraph with single maximal tree is a most complicated
part of the algorithm. Let us fix the spanning subgraph $R$ of
edges of a given color $\alpha$. We consider the flips from Lemmas
\ref{f8} and \ref{f9}. The flips change $R$.  According to the
Lemmas, after at most $3d$ steps either the number of edges in the
cycles is growing or there exists a single maximal tree.

The subroutine of pseudo code \large Flips \normalsize(spanning
subgraph $F$) returns either a stable pair or enlarges the number
of edges in cycles of the spanning subgraph. The subroutine uses
linear subroutines \large{FindParameters}\normalsize,
\large{MaximalTreeToStablePair} \normalsize and also has linear
time complexity $O(nd)$.

 We repeat the procedure with pseudo code \large Flips  \normalsize
 for a new graph if the number of edges in cycles after the flips grows.
 In the opposite case, we find a stable pair and then a homomorphic image
of a smaller size. For a graph of given size, the complexity of
this step is quadratic.

\subsection{Main procedure and complexity}
The Procedure \large{Main} \normalsize uses all above-mentioned
linear procedures and returns a synchronizing coloring (if exists)
 of the graph.
\\
\\
\large{Main}\normalsize()

1 arbitrary coloring of $G$

2  {\bf if} False(\large{CheckSinkSCC}\normalsize(graph $G$))

3 \quad {\bf return} False

4 {\bf if} \large{FindLoopColoring}\normalsize(F=SCC of $G$)

5 \quad {\bf return}

6  {\bf if} False(\large{FindGCDofCycles}\normalsize($SCC F$))

7 \quad {\bf return} False

8 \quad {\bf while} $|G|>1$

9 \qquad  {\bf if}\large{FindLoopColoring}\normalsize(F)

10  \qquad   \quad change the coloring of generic graph $G$

11  \qquad  \quad {\bf return}

12 \qquad {\bf for} every letter $\beta$

13 \quad \qquad {\bf do}

14 \qquad \qquad  {\bf if}
\large{FindTwoIncomingBunches}\normalsize(spanning subgraph,stable
pair)

15 \qquad \qquad \quad
\large{HomonorphicImage}\normalsize(automaton $A$,stable pair,new
A)

16 \qquad \qquad \quad \large{FindParameters} \normalsize($A =$
new $A$)

17 \qquad \qquad \quad {\bf break}

18 \qquad \qquad {\bf while} \large{Flips}\normalsize(spanning
subgraph $F$ of color ${\beta}$) = GROWS

19 \qquad \qquad \quad $F =$ new $F$

20 \qquad \qquad  \quad {\bf if}
\large{FindTwoIncomingBunches}\normalsize( $F$,stable pair)

21 \qquad \qquad \qquad
\large{HomonorphicImage}\normalsize(automaton $A$,stable pair,new
A)

22 \qquad \qquad \qquad \large{FindParameters} \normalsize($A =$
new $A$)

23 \qquad \qquad \qquad  {\bf break}

24 \qquad \qquad \large{MaximalTreeToStablePair} \normalsize
(subgraph, stable pair)

25 \qquad \qquad \large{HomonorphicImage}\normalsize(automaton
$A$,stable pair,new A)

26 \qquad \qquad \large{FindParameters} \normalsize($A =$ new $A$)

27 change the coloring of $G$ on the base of the last homomorphic
image
\\
\\
 Some of above-mentioned linear subroutines are included in
cycles on $n$ and $d$, sometimes twice on $n$. So the upper bound
of the time complexity is $O(n^3d)$.

Nevertheless, the overall complexity of the algorithm in a
majority of cases is $O(n^2d)$. The upper bound $O(n^3d)$ of the
time complexity is reached only if the number of edges in the
cycles grows slowly, the size of the automaton decreases also
slowly, loops do not appear and the case of two ingoing bunches
emerges rarely (the worst case). The space complexity is
quadratic.

 \end{document}